\documentstyle[12pt]{article}

 \newcommand \be {\begin{equation}}
\newcommand \bea {\begin{eqnarray} \nonumber }
\newcommand \ee {\end{equation}}
\newcommand \eea {\end{eqnarray}}
 
 \newcommand \s {\sigma}
 \newcommand \al {\alpha}

\begin{document}

\title{Attractor Neural Networks}
\author{Giorgio Parisi \\ Dipartimento di Fisica \\
Universit\`a di Roma {\it La Sapienza} }
\maketitle

\begin{abstract}

In this lecture I will present some models of neural networks that
have been developed in the recent years. The aim is to construct
neural networks  which work as associative memories.  Different
attractors of the network will be identified as different internal
representations of different objects. At the end of the lecture I
will present a comparison among the theoretical results and some of
the experiments done on real mammal brains.
\end{abstract}

\newpage

\section {Introduction}

The field of neural networks has been intensively studied by many
people coming from quite different disciplines: mathematics, computer
sciences, physics, biology and psychology. Each of them carries with
him his training, the methodology of his discipline and the taste for
what he believes it is important or interesting.

The goals of all these people are quite different. Roughly speaking
they can be divided into those who want to understand how the real
mammal brain works and those who want to use the information they can
gather in the study of the brain in order to build new brands of
computers. These new computers should behave in a more intelligent
way, with more good sense compared to the existing ones.

There are also others who are interested in understanding how complex
collective behaviour emerges from the interaction of many simpler
elementary objects. This is quite ageneral issue. Everywhere in
biology it can  be  observed that an assembly of objects behaves
quite differently from the isolated objects  and this is a crucial
phenomenon. Examples of emergent collective behaviour are
macromolecules from atoms, cells from  macromolecules, pluricellular
animals from single  cells,  species, evolution and ecosystems from
animals.

The task of deducing the laws for the higher levels from those  of
the lower level is not simple at all. Many physicists (and myself in
particular \cite{myself}) are fascinated  by this last question, also
because in the past we have already successfully answered to similar
questions, albeit in a much simpler  context. The emergence of
collective behaviour is quite familiar to us under the name of phase
transition. Many examples can be found: an assembly of spins becomes
ferromagnetic decreasing the temperature (or increasing it, if we
follow  the original notation by Celsius), a system of many electrons
behaves as superconductor...

Ferromagnetism, superconductivity are collective phenomena that can
be understood only
using the appropriate tools of statistical mechanics. For many
systems, especially for
superconductivity, the understanding of the behaviour of the whole
system from the
microscopic components and the construction of the corresponding
theoretical framework has
not been an easy job. Much more elaborate  constructions  are needed
in order to understand
some peculiar behaviour of disordered systems, like spin glasses,
where many microscopically
different, but macroscopically equivalent equilibrium states are
present \cite{mpv},
\cite{parisibook2}.

Nowadays physicists have succeeded in reaching a good control of many
of the observed
collective phenomena in physical systems and I believe that the task
of deriving macroscopic
laws form the microscopic ones in biological setting is our duty. It
seems to me that it is
nececessary to use  the techniques of statistical mechanics in order
to reach an  understanding of
the dynamics  and of the mutual interaction (also with DNA) of the
three thousands different
enzymes in an {\it  Escherichia Coli}. The times seem to  be ripe
for this task also  because in the last ten years we have started to
understand the behaviour of
physical complex  systems, mainly disordered systems as spin glasses,
which display many different
equilibrium  states. Our toolbox of statistical mechanics contains
rather sharp and sophisticated
instruments.

It approach to biology physicists have often reproached for
oversimplifying. I believe that
physicists do oversimplify (using the standards of biologists) and
they do it for good reasons.
Let me present the two most important reasons:
\begin{itemize}
  \item
 Modern physics start at Galileo's times with the seemly unnatural
choice of neglecting
friction.
In every day life friction dominates; the force is proportional to
the speed as in the old
approach of Aristotle, not to the acceleration. Without friction we
are unable even to walk, as
it
can be verified when we find ice on the road. A world without
friction would not look like the
real world.

However physics started by first extrapolating from the experiments
the properties of a
frictionless world and by constructing a theoretical framework able
do describe this artificial
world \footnote{Let me quote from the introduction from a book of
Tartaglia (freely translated
into modern language): {\it In this book
we will study those objects that move with constant speed in the
horizontal direction and with
constant acceleration in the vertical direction. If real cannon balls
do no follow this law, the
worse for them, we will not speak about them.}}, which culminated in
NewtonÕs Mechanics.
Friction has been included in the theory at a much later stage.

Many times in the history of physics progress have been achieved by
selecting
some features that are supposed to be simpler to analyse and  by
neglecting the rest for the time
being, up to the moment in which the theory was able to cope  with
new features  and
nowadays this approach is an  automatic reflex of physicists. The
great simplifications (which we
will see later) introduced  by
Hopfield in its study of Attractor Neural Networks are a magnificent
step backward, in the
tradition of physics.
\item
There is also a more technical reason for oversimplifying, which has
been discovered in the
study of second order phase transitions by Kadanoff and Wilson. It
carries the name of {\it
universality}. The basic idea is simple. The number of qualitatively
different collective
behaviours of a system at the macroscopic level is not very high and
it is definitively much less
that the number of different microscopic systems. Thousands (or
millions) of different
ferromagnetic materials (of Ising type) exist and in all of them the
specific heat  diverges
at the phase transition point as $|T-T_c|^\alpha$. The value of
$\alpha$ (about
$-.11$) is the same for all the systems. The qualitative and
quantitative properties of the
collective behaviour
depend on the kind of collective behaviour that is observed, not on
the microscopic details of
the system.

Of course quite different system may have qualitative different phase
transitions (some
become ferromagnetic, others superfluids or glasses). Phase
transitions may be divided in a
few  universality classes (at least for homogeneous systems)
\footnote{ In the same spirit of
Thom's classification of catastrophes}; the qualitative behaviour and
some of the quantitative
features near the phase transition point are constant inside a given
universality class and
they do not change with small change of the systems. In other words
the collective behaviour
can be understood in self consistent way without making reference to
the microscopic
dynamics. This program reached its higher success for two dimensional
systems, where all
possible collective behaviour of systems belonging to a quite large
class has been classified
using simple consistency requirements for the collective behaviour.

If the understanding of the collective behaviour can be obtained by
studying the universality
classes and not a particular system, quite different systems that are
supposed to belong to the
same universality class are equivalent as far as the collective
behaviour is concerned. One of
the first reaction of a theoretical physicist to a new problem is to
search the
simplest system that should belong to the same universality class.
Universality suggests that
unwanted complications can be neglected and often we do not need to
reintroduce them at a
later
stage, because they do not change the main results. Of course a large
amount of experience
and maybe of Serendipity is needed to discriminated the unwanted
complications from the
essential properties.
   \end{itemize}

In this lecture I will describe some of the developments than have
been done in the recent
years on the construction of a network of neurons which should behave
as associative memory.
Associative memory is memory in the usual human sense\footnote{The
world associative is
used to distinguish this kind of memory by the computer memory
(addressable memory) where each address identifies the content of the
address.}: we remember
names, faces, notions,  we recognise known objects and we are able to
recall them also if we
receive only a partial  information . At the end of the lecture I
will present a
comparison of the predictions of the models with experiments with
real mammal brains.

\section {A Schematic Description of Neurons}

Before entering it the details of the model it may be convenient to
present a schematic
description of the neurons and their mutual interactions\cite{BS}. We
will mainly consider the
electric  activity of the neuron, also because it is the easiest to
measure. If we consider the
difference  in
voltage of the neuron with the surrounding liquid, we observe fast
spikes of about 100 mV of
amplitude of a time duration of about a millisecond. The number of
spikes per second may
range from 1 to a few hundreds, although in normal physiological
conditions is not greater
than a few tens.

The activity of a neuron depends on the signal it receives from other
neurons. Indeed
different neurons are connected by synapses, which are unidirectional
devices needed to
transmit information. If the neuron $A$ has a synapse toward $B$,
when $A$ is active,
neurotransmitters are released by $A$ and absorbed by $B$. The
release of neurotransmitters
increases when the activity of the neuron increases.

The synapses may be excitatory or
inhibitory. The neurotransmitters increases the activity of $B$ if
the synapse is excitatory,
and they decreases the activity if the synapse is inhibitory. A given
neuron may receive
signals from many neurons (the number of neurons is $10^4$ in
average, with a range which goes from
2 to $10^5$). The  activity of a neuron will be high if the
excitatory signals overcame the
inhibitory ones,  otherwise it would be low.

Different synapses will have different effects. The efficacy of a
given synapse depends on
time and it is usually believed that there is some physiological
mechanism
which strengthen those synapses in which both neurons are
simultaneously active (the Hebb's
rule). The details  of this mechanism in vertebrates are unclear at
the present moment, although
a large  amount of information  has been collected.

Synapses allow a neuron to send a message specifically to an other
neuron, which may be quite
far away. Neurons also send non specific messages to all nearby
neurons by producing
neurotransmitters which diffuse in the intercellular liquid. Here we
are not going to consider
this activity.

If we try do construct a model for a set of $N$ interconnected
neurons (i.e. a neural network),
we can parametrize the activity of the neurons by $N$ functions
$\sigma_i(t)$, $i=1,...,N$, and
the synaptic strength by $N^2$ functions $J_{i,k}(t)$.

The stimulus of the network on a given neuron ($i$) is assumed to be
given by
\be
S_i(t)=\sum_{k=1,N} J_{i,k}(t) \sigma_k(t),
\ee
i.e. the total stimulus is simply the sum of the stimulus coming form
each neuron, which we
assume to be equal to the product of the synaptic strength with the
activity of the neuron
producing the stimulus
\footnote {In writing eq. (1) we have neglected the propagation time
of the stimulus. Adding a
variability in the propagation times leads to the appearance of new
phenomena.}.

The dynamic equations for the neuron are supposed to be in the
simplest case:
\be
{d\sigma_i \over dt}= f(\sigma_i,S_i),
\ee
where $f$ is a non linear function of its arguments.

In order to fully describe the network, expecially in the learning
stage, we should also give the
equations which control the time  evolution of the synaptic strengths
($J$), however for reasons
of time I will not discuss this  point, which is very interesting and
unfortunately not so well
understood. In  the following we will assume that the
 dynamics of the $J$'s has been such to produce synaptic couplings
with of the kind we
postulate.

\section{Associative memory}

It is believed that an internal representation (i.e. a pattern of
neuron activities) is associated
to
each object or category that we are able to recognise and remember.
It is also believed that an
object is memorised by changing the synaptic strengths in an
appropriate way.
In order to have an associative memory  the
dynamics of the neurons should such that
\begin{itemize}
\item
If the neurons are set equal to $\mu_i$ at time 0, where the $\mu_i$
correspond to one of  the
learned pattern, the neuron activities  $\sigma_i(t)$ should remain
for all times not too far
from $\mu_i$. In other words the learned pattern correspond to a
stable  attractor points, in a
statistical sense (i.e. small fluctuations around the fixed points
are allowed).

\item
If we apply an external stimulus which push the neurons in the
direction of being equal to
$\tau_i$ , and if the $\tau_i$ only partially correspond to one of
the learned patterns $\mu_i$,
the neurons should evolve
toward the $\mu_i$.
 This property implies that the memory is able to retrieve the
information on the whole object
from
the knowledge of a part of it or in presence of partially wrong
information.

\item
If the $\tau_i$ of the external stimulus are very different from the
$\mu_i$ of any stored
pattern, the
system should recognised the absence of a match. This goal may be
obtained if in this case no
fixed
point is reached  and a chaotic behaviour is observed. In this case
the memory is called
cognitive because it knows when a correct result is obtained:
reaching an
attractor (a fixed point) has the meaning of recalling the
information, while no stable point
implies the absence of
recognition.
\end{itemize}

It is clear that physiological constraint implies that many different
patterns of activities may
be stored same neural network. Our model of network should be such
that many different
attractors are present. From this point of view memory is very
similar to a spin glass, where
many different equilibrium states are possible\cite{mpv,parisibook2}.

\section{The Hopfield model}

The Hopfield model is the simplest version of a model in which many
different attractors do
exist \cite{HOP}. In this model the neuron activities are
conventionally taken to be $\pm 1$ and
the  time $t$ is also an integer valued quantity. This choice is
rather unnatural; the choice
$\s=0$ or  1,
would sound much better, however no crucial qualitatively difference
are among these two
cases, as we shall see later.

One step (i.e. the increase of $t$ by an unity) for the dynamics of
the neuron activities is
obtained by applying (for all $i$) the rule
\be
\sigma_i(t+{i+1 \over N} )= {\rm sign}(S_i(t+{i \over N})),
\ee
where we still have
\be
S_i(t)=\sum_{k=1,N} J_{i,k}  \sigma_k(t).
\ee

If the matrix $J$ is symmetric ($J_{i,k}=J_{k,i}$), this dynamics
corresponds to a sequential
algorithm for looking for the
minimum of the Hamiltonian
\be
H= -\sum_{i,k=1,N} J_{i,k}  \sigma_i\sigma_k.
\ee

The main advantage of the Hopfield model was that at this stage the
model is very similar to the
zero temperature dynamics of a statistical mechanics model. The
analogy becomes more strong is we
introduce the possibility of errors and we write
\begin{eqnarray}
\sigma_i(t+{i+1 \over N} )&= {\rm sign}(S_i(t+{i \over N}))
 {\rm  \  with \ probability} & |\tanh (S_i((t+{i \over N})/T)|, \\
\nonumber
\sigma_i(t+{i+1 \over N} )&= -{\rm sign}(S_i(t+{i \over N}))
 {\rm \  with \ probability} & 1-|\tanh (S_i((t+{i \over N})/T)|.
\eea
 In this case we obtain the dynamics of a statistical mechanics
system at temperature $T$ with the
random Hamiltonian (5), which can successfully studied using the
techniques developed for spin
glasses. In the limit $T \to 0$, we recover the original error free
dynamics of eq. (3).

 It is possible to prove that for the original zero temperature
dynamics a fixed point is reached for large time, i.e. the  $\s$'s
become time independent. Indeed
the Hamiltonian decreases every times that a $\s$  changes its value.
The absence of limiting
cycles is a very strong simplification and it is the  main reason for
considering this model
\footnote
{We could also allow the possibility of having a periodic
behaviour with period 2, i.e. the $\s$  becomes time independent
separately for odd and even
times. In this case we should update the neurons in a sequential way,
i.e. by setting
$\sigma_i(t+1)= {\rm sign}(S_i(t)$. The properties of this model are
rather similar to the one
described in the text and it is possible that only fixed points or
limiting cycles of period two
may be present. }.

Our aim is to store in network $M = \alpha N$ different patterns,
which are denoted by
$\mu_i^p$, where the index $i$ runs from $1$ to $N$ and the index $p$
runs  from $1$ to
$M$.  The matrix $J$ is assumed to have the following symmetric form
\be
J_{i,k}= \sum_{p=1,M} \mu_i^p\mu_k^p,\label{HEBB}
\ee
which has a strong Hebbian flavour. Indeed it correspond to say that
the matrix $J$ changes
by the quantity $\mu_i^p\mu_k^p$ each time that the network learns
the new pattern $\mu_i^p$.
The symmetry of the matrix $J$, although is not physiologically
realist, is necessary for using
the statistical mechanics techniques and for being able to consider
the evolution of the network
as a process of energy minimisation.

 We can ask the question if there are configurations of the
$\s$ which are fixed point of the  dynamics, i.e. they satisfy the
equations
 \be
\s_i= {\rm sign} (\sum_{k=1,N} J_{i,k} \s_k)
\ee
and are not far from one of the $\mu^p$'s for some value of $p$.

This requirement may be
stated in a more precise way. We can define the distance among a
configuration $\s$ and a configuration $\tau$ as
\be
d(\s,\tau)=1/N\sum_{i=1,N} (\s_i-\tau_i)^2/4.
\ee
In other words $d(\s,\tau)$ is the percentage of neurons which are
different in the two
configurations. If the two configurations are uncorrelated, the
distance is .5.

We look for existence of configurations $\s$, which satisfy eq. (8)
and have a small distance
with one of the  $\mu$'s.  This problem has been carefully studied in
the limit where $N$ goes
to infinity\cite{Amit}.
 In the simplest case we assume that the different $\mu$'s are
uncorrelated one
from the other and they have in average the same number of +1 and -1.
In this situation one
finds the existence of two regimes:
\begin{itemize}
\item For $\alpha < \alpha_c$, where $\al_c \approx .14$, one can
find at least a
configuration $\s$ very near to each of the $M$ configurations
$\mu^p$. More precisely we can
find it at a distance less that $d_{Max}(\al)$,  where $d_{Max}(\al)$
is
an increasing function of $\al$, which reaches its maximum (i.e.
$0.01$) in this interval  at
$\al_c$. The number of
these solutions is increasing with $N$, but very slowly (i.e. as
$\exp(\omega N)$ with a very
small value of $\omega$).
\item For $\al > \al_c$ no solutions exist with small distance and
one find solution only at
distances of order of $.4 - .5$.
\end{itemize}
In the first regime the network works as an associative memory,
however, when the loading
become greater than $\a>a_c$, the memory stops to work and goes into
a state of total
confusion. This first order transition is a feature
of the model which disappear changing some of the characteristics of
the model. In all models
there exist a critical value of $\al$ such that the system does not
work as a memory for
$\al>\al_c$.

\section{More refined models}

The Hopfield model contains many unrealistic features. Many more
refined models have been
proposed. The restriction to the neuron activity to be $\pm 1$ can be
easily removed without
dramatic consequences. Different choices of the probability
distribution of the patterns and of
the learning procedure have been studied. Other features  can be
added, e.g. the
dilution of the network, which consists in assuming that not all the
neurons are connected and
some of the $J$'s are strictly zero.

 I cannot mention these interesting
results for lack of time. I  will only recall a very interesting
result  obtained by the late
Elisabeth Gardner \cite{Gardner}.  Here we do not commit ourselves to
any learning algorithm and
we  can ask if we can find a matrix $J$ such that all the patterns
are fixed point of the
dynamics. A  very interesting computation show that when the patterns
are uncorrelated, in the
limit $N \to  \infty$ this is always possible for
$\al<1$ and never possible for $\al>1$. The interest in this kind of
results is that they tell us
what a network can do independently of learning algorithm, i.e. of
the dynamics for the $J$Õs.

 Using a symmetric form for the matrix $J$ we remain in the framework
of
equilibrium statistical mechanics.
New qualitative features appear if the $J$ are asymmetric \cite{P}.
Indeed in the Hopfield model
the
evolution of the $\s_i$ is such that a fixed point is always reached,
also if this fixed point has
nothing to do with one of the stored patterns. If we introduce
asymmetries in the matrix $J$, it
can be shown that the evolution does not leads always to a fixed
point and the spurious fixed
points not correlated with the
stored patterns do disappear. If the network does not reach a fixed
point near one of the stored
pattern, the time evolution of the network is chaotic and the
variables $\s_i$ will change value
in an apparent random way. This results is very interesting because
one obtains without effort
a good feature (the possibility of chaotic states) by removing one of
the artificial
characteristic of the model (i.e. the symmetry of the $J$'s).

A very simple model in which this effect can be studied is the
asymmetrically diluted model, where
\be
J_{i,k}=J^H_{i,k} A_{i,k},
\ee
$J^H_{i,k}$ being given by eq.(\ref{HEBB}) and $A_{i,k}$ is an
asymmetric matrix, whose elements
are  1
with probability $z$ and 0 with probability $1-z$.

In the limit of small $z$ the model can be easily studied
analytically and after simple computations
one  finds
that the storage capacity $\al_c$ is equal to ${2z / \pi}$ and that
the large time behaviour
of  the network is fully chaotic when it is not near one of the
attractors \cite{DGZ}.

A very large amount of work has been done in these recent years also
on many other problems (e.g.
learning of rules from example)  \cite {SARA,ST,P2,OPPER} and in most
of the cases
the extension of the techniques developed for statistical mechanical
systems has been very
fruitful.

\section {Real experiments}

We have seen that we can construct many model with different features
which work as
associative memories. It is quite likely that these results will be
eventually important for
the projects  of new generations of computers. At the present moment
a
very interesting practical application is
to use neural networks as fast triggers in order in high energy
experiments. i.e. at super
colliders. In this case one should decide if an event is interesting
or not in a rather short time,
of the order of $10$ nanoseconds \cite{BOOK}. Here a neural  network
may be able to perform this
rather complex task in such a short time using its massive
parallelism, i.e. using many neurons,
each of them being physically implemented in Silicium.

{}From a biological point of view we are interested  to known if
attractors do really exist in
mammal
brains and how the brain perform very complex task (for example it is
able to recognise if different
image do  correspond to the same physical object). In order to
address to this problem we have
to identified experimentally the memory cell in which the information
is stored.

 In the past many neurons have been found which respond (i.e. they
switch from a low
firing rate to an higher one) in a selective way to a given external
stimulus (for example
visual).
The stimulus was sometimes highly complex (e.g. a face), sometimes
very simple (an horizontal
line). Unfortunately  up to a recent time  the activity of the
neurons
disappeared when the stimulus was removed.

Very recently neurons have been found which remain active in a
selective way after the
removal of the stimulus for a period larger that 16 seconds
\cite{M1,M2}. In the experiments
monkeys, of  the
species {\it Macaca Refus}, were trained to learn and to memorise 100
fractal patterns, which were
presented in a given order.

After learning  these patterns were presented to the monkeys in pairs
at a time distance of
about
20 seconds. They were supposed to act differently if the first
pattern was equal to the first
and a
fruit juice was given to them if they performed correctly. It was
found that in small area of
about 1 mm$^3$, there are neurons which respond in selective way to
the visual presentation of
only few of the  patterns.

It is reasonable to suppose that the recorded activities come from
cells that belong to the
attractor, not from cells that are excited by the attractor. If this
hypothesis is correct, it
should be possible to start an experimental study of the property of
the attractor. Many
interesting features of the experiment have been compared with
theoretical predications.

A very interesting and unexpected feature of the experiment is the
existence of correlations
in the neural activities of different patterns. The patterns have
been presented in a fixed,
randomly chosen order. The patterns itself are uncorrelated, however
the probability that a given
neuron is active after the presentation of the $i^{th}$ pattern is
much higher that the
average, if the neuron is active after the presentation of $k^{th}$
pattern  and $|i-k|$ is not
large (e.g. smaller that 6). The neural activities corresponding to
patterns which have been
learned at nearby times are correlated.

It has been shown \cite{ABT} that these correlations arise naturally
in a relative simple
model of neural network and that the experimental features of the
correlations is in good
agreement with the theoretical predictions.

It is quite likely that in the future experiments and theory will
evolve in a convergent way and
it will be possible to obtain a much more detailed confirmation (or
refutation) of the
theoretical approach presented here.
\vskip 12pt
I hope to have given to the audience some flavour of the large amount
of work that has been done
in the field of neural network in the last years. Much more could
have been said, but the time is
over.

\end{document}